\documentclass[prl,twocolumn,notitlepage,superscriptaddress,longbibliography]{revtex4-1}
\usepackage{amsmath}
\usepackage{amssymb}

\usepackage[unicode=true,colorlinks=true,citecolor=blue,urlcolor=blue]{hyperref}

\usepackage{xpatch}

\makeatletter
\patchcmd{\@ssect@ltx}
    {\addcontentsline{toc}{#1}{\protect\numberline{}#8}}
    {}
    {}
    {}
\makeatother %remove unnumbered sections from top

\usepackage{bm}
\usepackage{epsfig}

\usepackage[normalem]{ulem}

\newcommand {\diag}{\mathop{\mathrm{diag}}\nolimits}

\renewcommand {\Re}{\mathop\mathrm{Re}\nolimits}

\renewcommand {\phi}{{\varphi}}
\newcommand {\rmi}{{\rm i}}
\newcommand {\rmd}{{\rm d}}

\newcommand {\e}{{\rm e}}
\newcommand {\eps}{\varepsilon}

\begin{document}
\title{Quantum Hall phase emerging in an array of atoms interacting with photons
}
\author{Alexander V. Poshakinskiy}
\affiliation{Ioffe Institute, St. Petersburg 194021, Russia}

\author{Janet Zhong}
\affiliation{Nonlinear Physics Centre, Australian National University, Canberra ACT 2601, Australia}
\author{Yongguan Ke}
\affiliation{Nonlinear Physics Centre, Australian National University, Canberra ACT 2601, Australia}
\affiliation{Guangdong Provincial Key Laboratory of Quantum Metrology and Sensing $\&$ School of Physics and Astronomy, Sun Yat-Sen University (Zhuhai Campus), Zhuhai 519082, China}
\author{Nikita A. Olekhno}
\affiliation{ITMO University, St. Petersburg 197101, Russia}
\author{Chaohong Lee}
\affiliation{Guangdong Provincial Key Laboratory of Quantum Metrology and Sensing $\&$ School of Physics and Astronomy, Sun Yat-Sen University (Zhuhai Campus), Zhuhai 519082, China}
\affiliation{State Key Laboratory of Optoelectronic Materials and Technologies, Sun Yat-Sen University (Guangzhou Campus), Guangzhou 510275, China}

\author{Yuri S. Kivshar}
\email{yuri.kivshar@anu.edu.au}
\affiliation{Nonlinear Physics Centre, Australian National University, Canberra ACT 2601, Australia}
\affiliation{ITMO University, St. Petersburg 197101, Russia}

\author{Alexander N. Poddubny}
\email{poddubny@coherent.ioffe.ru}
\affiliation{Ioffe Institute, St. Petersburg 194021, Russia}
\affiliation{Nonlinear Physics Centre, Australian National University, Canberra ACT 2601, Australia}
\affiliation{ITMO University, St. Petersburg 197101, Russia}

\begin{abstract}
Topological quantum phases underpin many concepts of modern physics. While the existence of disorder-immune topological edge states of electrons usually requires  magnetic fields, direct effects of magnetic field on light are very weak. As a result, demonstrations of topological states of photons employ synthetic fields engineered in special complex structures or external time-dependent modulations. Here, we reveal that the quantum Hall phase with topological edge states, spectral Landau levels and Hofstadter butterfly can emerge in a simple quantum system,
where topological order arises solely from interactions without any fine-tuning. Such systems, arrays of two-level atoms (qubits) coupled to light being described by the classical Dicke model, have recently been realized in experiments with cold atoms and superconducting qubits.  We believe that our finding will open new horizons in several disciplines including quantum physics, many-body physics, and nonlinear topological photonics, and it will set an important reference point for experiments on qubit arrays and quantum simulators.
\end{abstract}

\date{\today}

\maketitle

The study of electrons propagating in magnetic fields has been driving many problems of physics since the discoveries of the Landau levels~\cite{ezawa2013quantum} and a self-similar structure of the energy spectrum in crystals subjected to ultra-high magnetic fields~\cite{Azbel1964, Hofstadter1976}. Quantum Hall effect~\cite{TKNN1982} and topological insulators~\cite{Bernevig2006,Hasan2010} brought the concepts of topological phases to condensed matter physics. However, many effects predicted long time ago including the Hofstadter butterfly spectrum have been realized only recently~\cite{Dean2013}.

These developments inspired a rapid progress in {\em topological photonics} aiming at creating robust edge states of light immune to disorder~\cite{Ozawa2019,wang2009,Haldane2008a,Chang2020}. Since the effects of magnetic fields on light are weak, the realisation
of topological concepts in photonics requires artificial structures and metamaterials~\cite{Khanikaev2017}. Alternative approaches rely on  time modulation of structure parameters~\cite{hauke2012,Sounas2017,Roushan2017,Dutt2019} or engineered nonlinearities~\cite{Hadad2018,smirnova2019nonlinear}. These approaches allow creating effective gauge fields in real or synthetic dimensions, and mimick the effects of magnetic fields or spin-orbit couplings for photons.

%%%%%%%%%%%%%%
\begin{figure*}[t]
\includegraphics[width=0.7\textwidth]{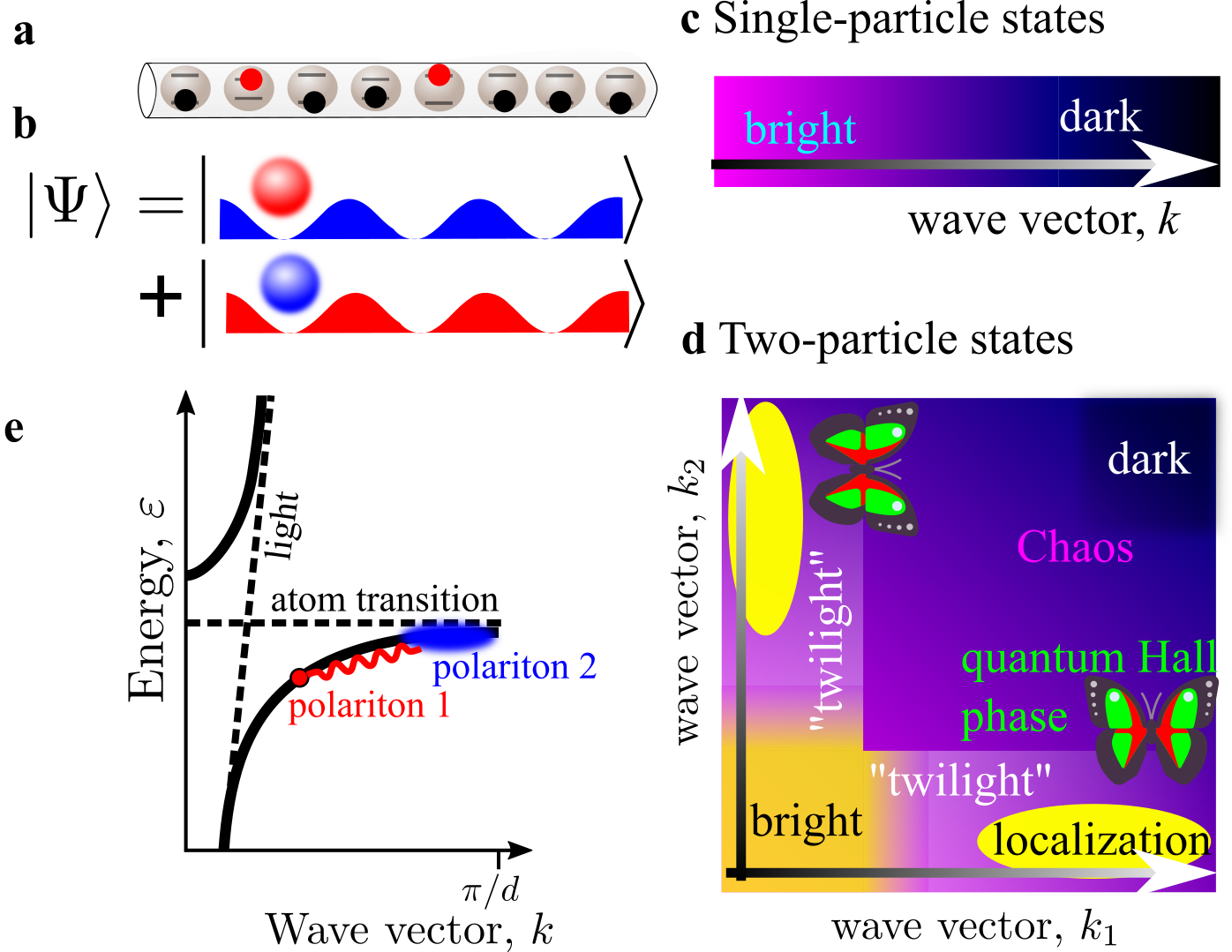}
\caption{{\bf Emergence of quantum phases due to polariton-polariton interaction.} {\bf (a)} Double-excited array of two-level atoms (qubits) in a waveguide. {\bf (b)} Two-polariton quantum states where each indistinguishable polariton induces a potential for the other one.  {\bf (c)} and {\bf (d)}: Classification of single- and double- excited states of the atomic array depending on the wave vector of the excitations.  Butterflies in {\bf (d)} indicate the regions  where the quantum Hall phase and Hofstadter-like butterfly spectrum emerge from the interaction of two excitations. Regions of ``twilight'' states~\cite{Ke2019} and interaction-induced localization~\cite{Zhong2020} are also shown. {\bf (e)} Single-particle polaritonic dispersion. Interaction of a lower-branch polariton with small $k$ and that with large $k$ is illustrated.
}\label{fig:1}
\end{figure*}
%%%%%%%%%%%%%%

Here, we uncover that the hallmarks of the quantum Hall phases, including  Landau energy levels,  topological edge states, and Hofstadter butterfly spectrum, can appear in a simple quantum system:  an array of closely spaced two-level atoms (qubits) coupled to photons in a waveguide, see Fig.~\ref{fig:1}a.  In this system, photons become strongly coupled to atoms and create {\em polaritons}. These polaritons are not independent but strongly interacting, because one atom cannot absorb two photons simultaneously~\cite{Birnbaum2005}. While the considered model is paradigmatic for quantum optics~\cite{Roy2017,KimbleRMP2018,FriskKockum2019}, its  two-particle Hilbert space was not analyzed until recently. As shown in Figs.~\ref{fig:1}(c,d), when the polariton wave vector is comparable with that of light, a collective  atomic state is easily excited optically, and it generally gets ``darker" for larger wave vectors. In a two-particle ``bright" state, the wave vectors of both excitations are small, which corresponds to the Dicke superradiance~\cite{Dicke1954}. Novel two-particle dark states, where both wave vectors are large, were predicted only last year,  and they originate from fermionization of strongly interacting polaritons~\cite{Molmer2019}.
It has also been suggested that interactions in the  corner regions \cite{Ke2019} of the diagram of Fig.~\ref{fig:1}d can  localize one of the two polaritons  in the  center of the array~\cite{Zhong2020}.

In this paper, we predict novel types of topological edge states driven by polariton-polariton interactions in the regions indicated  by butterflies in Fig.~\ref{fig:1}d. Here, one polariton forms a standing wave with multiple nodes and a periodic potential for the other indistinguishable polariton, see Figs.~\ref{fig:1}(b,e). As a result, the interaction is described by the {\it  self-induced} Aubry-Andr\'e-Harper~\cite{aubrey} model that is mathematically equivalent to the quantum Hall problem on a lattice~\cite{Kraus2012,Poshakinskiy2014}.

The striking novelty of our prediction is that the quantum Hall phase can emerge for interacting  indistinguishable  particles  without any special fine-tuning. The periodic modulation is an intrinsic feature that arises naturally due to the polariton-polariton  interactions, in a sharp contrast to previous studies~\cite{Baboux2017,Ke2020}, where one had to impose the modulation deliberately, either by engineering the lattice~\cite{verbin2013,Ke2020} or applying external fields~\cite{Roushan2017,Chang2020}. The full Hofstadter butterfly-like spectrum could be obtained in a single shot from just one fixed atomic array, eliminating the need to continuously tune an external magnetic
field in a conventional setup~\cite{Roushan2017,Dean2013}. Our results apply to the experiments with cold atoms~\cite{Corzo2019} or superconducting qubits coupled to a waveguide~\cite{Astafiev2010,Wang2019,Haroche2020,Browaeys2020,Blais2020,Clerk2020,Carusotto2020} and  emerging 	quantum simulators based on  excitonic polaritons~\cite{Ghosh2020}. This offers new possibilities to understand quantum many-body topological phases of interacting matter and protect them against decoherence.

\begin{figure*}[t]
\includegraphics[width=\textwidth]{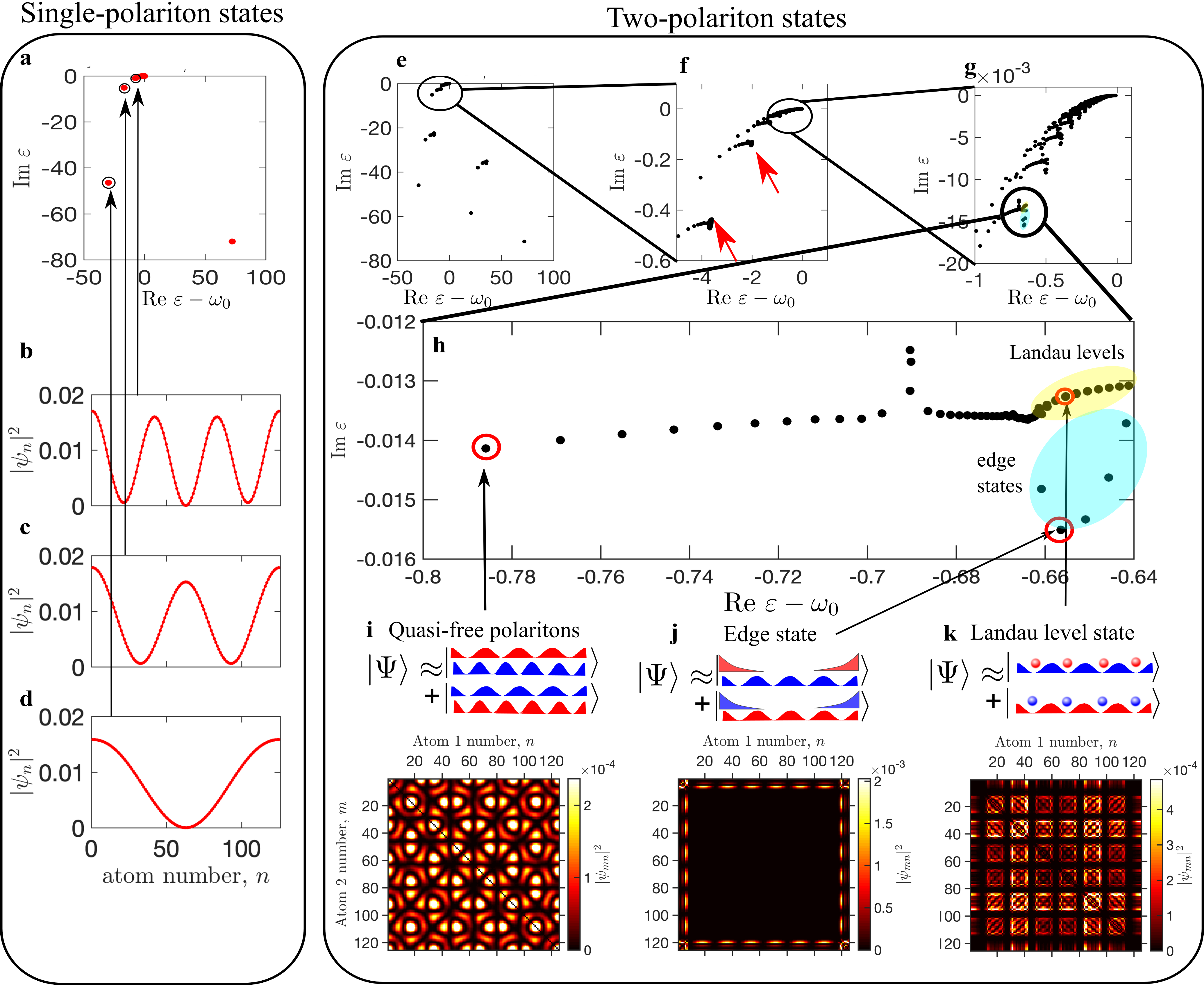}
\caption{{\bf Single- and two-polariton energy spectra}.
{\bf (a)}: complex energy spectrum of single-polariton modes.  Three characteristic eigenstates are shown in  panels
{\bf b}--{\bf d}.  {\bf (e--h)}: Two-polariton energy spectrum zoomed in different scales.
 {\bf (i,j,k)}:  Spatial color maps of different characteristic two-polariton eigenstates $|\psi_{nm}|^2$.  Calculation has been performed for $N=125$~atoms and $\omega_0d/c=0.02$. Energy is measured in units of $\Gamma_0$.
}\label{fig:2}
\end{figure*}
%%%%%%%%%%%%%%
%%%%%%%%%%%%%%%%%%%%%%%%%%%%%%%%%%%%
\section*{Two-polariton states}
%%%%%%%%%%%%%%%%%%%%%%%%%%%%%%%%%%%%
%%%%%%%%%%%%%%
\begin{figure}[t]
\includegraphics[width=0.45\textwidth]{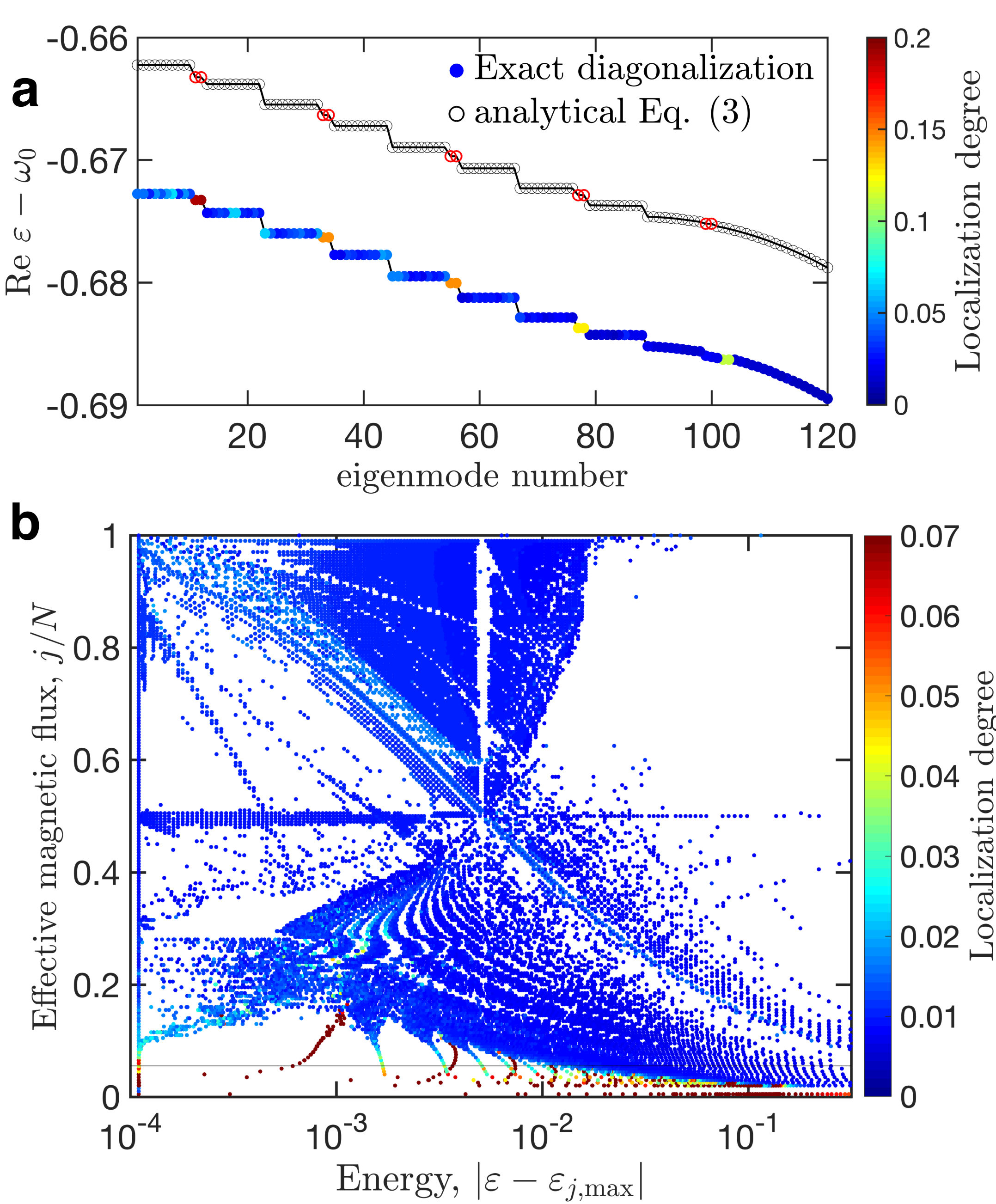}
\caption{{\bf Self-induced Hofstadter butterfly.}
{\bf (a)} Energy spectrum  for the two-polariton states in the cluster corresponding to $j=11$, calculated from the approximate Eq.~\eqref{eq:main} and by the exact diagonalization of the two-particle Hamiltonian Eq.~\eqref{eq:H}.
{\bf (b)} Butterfly energy spectrum obtained   by the exact diagonalization as a function of cluster index $j$, determining the effective magnetic field.
Localization degree is determined as the inverse participation ratio of the vector $\chi$ in Eq.~\eqref{eq:psi} and is shown by color.
Thin horizontal line in (b) indicates magnetic field $j/N=11/200$, corresponding to panel (a).
Calculation has been performed for $N=200$ and $\varphi=0.02$,
energy is measured in the units of $\Gamma_0$.
}\label{fig:3}
\end{figure}
%%%%%%%%%%%%%%
We consider a periodic array of two-level atoms (qubits) coupled to light, described by an effective Dicke-type Hamiltonian \cite{Molmer2019,Ke2019,Albrecht2019}
\begin{equation}\label{eq:H}
H=\sum_n\omega_0 \sigma_n^\dag\sigma_n^{\vphantom{\dag}}-\rmi\Gamma_0\sum\limits_{n,m}\e^{\rmi \omega_0 d|n-m|/c}\sigma_n^\dag\sigma_m^{\vphantom{\dag}}\:,
\end{equation}
where $\sigma_n^\dag$ is the operator creating excitation of the atom $n$ with the resonance frequency $\omega_0$, $(\sigma_n^{\dag})^2=0$ and $\Gamma_0$ is the radiative decay rate of a single atom. While for  $d=0$ the Hamiltonian \eqref{eq:H} is equivalent  to the conventional Dicke model~\cite{Dicke1954}, even small  interatomic spacings $0<d\ll 2\pi c/\omega_0$ make the model considerably richer.

Single-particle eigenstates of Eq.~\eqref{eq:H} are polaritons with the energy dispersion $\eps(k)=\omega_0+\Gamma_0/[\cos kd-\cos( \omega_0d/c)]$~\cite{Ivchenko1991,Albrecht2019}, schematically shown in Fig.~\ref{fig:1}e. The dispersion consists of two polaritonic branches, resulting from the avoided crossing of light with the atomic resonance and  we focus  on the lower   branch. In the finite array of $N$ atoms, the wave vectors are quantized $k_jd=\pi j/N$, $j=1,2,\ldots N$~\cite{vladimirova1998ru},  and the eigenstates are standing waves, see  Figs.~\ref{fig:2}{a--d}.   Negative  imaginary part of the energies in Fig.~\ref{fig:2}a characterizes radiative decay into the waveguide. Crucially, the spectrum in  Figs.~\ref{fig:2}a  condenses near the resonance $\eps=\omega_0$, where   the group velocity of polaritons decreases.

Next, we proceed to the  double-excited states $\Psi=\sum_{n,m}\psi_{nm}\sigma_n^\dag\sigma_m^\dag|0\rangle$.
Their spectrum,  obtained  from  the Schr\"odinger equation $H\Psi=2\eps\Psi$,   is shown in Figs.~\ref{fig:2}e--h in different energy scales and demonstrates a distinct   clustered structure. Each cluster resembles the single-particle  spectrum in Fig.~\ref{fig:2}a and  is  formed by a polariton with a certain wave vector $k_j$ interacting with  polaritons with larger wave vectors. Therefore, most of the spectrum in Figs.~\ref{fig:2}{e--g}  could be described by $\eps\approx (\eps_j+\eps_i)/2$, where $\eps_j$ and $\eps_i$ are the single-particle energies from
 Fig.~\ref{fig:2}a. However, the dense part of the cluster, which corresponds to $\eps_i \to \omega_0$ (see red arrows in Fig.~\ref{fig:2}f), is drastically transformed by the interaction. Three characteristic  states  from the cluster with $j=7$ are presented in Figs.~\ref{fig:2}i--k.  While the state  in Fig.~\ref{fig:2}i is just a symmetrized product of two   standing waves, weakly modified by interaction, the
role of the interaction dramatically increases for  $\Re \eps -\omega_0>-0.66\Gamma_0$  in Fig.~\ref{fig:2}h. The spectrum is   split by interaction into relatively delocalized states with smaller radiative decay rate (yellow ellipse in Fig.~\ref{fig:2}h and Fig.~\ref{fig:2}k) and the states with larger radiative losses, where one of the two polaritons is localized at the edge of the structure  (blue ellipse in Fig.~\ref{fig:2}h and Fig.~\ref{fig:2}j).

This interaction-induced transformation of the two-polariton spectrum  is our central result. The delocalized states are almost  $(j-1)$-fold degenerate, where $j$ is the cluster number  and correspond to the Landau levels in the effective magnetic field. The states in Fig.~\ref{fig:2}j come in degenerate pairs corresponding to topological edge states localized at the opposite sides of the array.

%%%%%%%%%%%%%%%%%%%%%%%%%%%%%%%%%%%%%%%%
\section*{ Landau levels, topological edge states, and Hofstadter butterfly}
%%%%%%%%%%%%%%%%%%%%%%%%%%%%%%%%%%%%%%%%
We now present  an analytical model  explaining  the topological origin behind  the interaction-induced edge states  in Fig.~\ref{fig:2}j. In the basis $|x\rangle=\frac1{\sqrt{N}}\sum_{n=1}^N\exp(\rmi \omega_0d|x-n|/c)\sigma_n^\dag|0\rangle$, $x=1,2, \ldots N$ \cite{Zhong2020,Poddubny2019quasiflat}, the following ansatz can be used for the  two-polariton state
\begin{equation}\label{eq:psi}
\psi_{xy} = \psi^{(j)}_{y} \chi_{x}+\psi^{(j)}_{x}\chi_{y}\:,\quad x,y=1\ldots N
\end{equation}
where $\psi^{(j)}_{x}$ and $\chi_{x}$ are the wave functions of the first and second polaritons. The former is assumed known and corresponds to the standing wave $\psi^{(j)}_{x} = \cos k_j(x-\tfrac{1}{2})$. To determine the latter, we derive the Schr\"odinger equation, that accounts for interaction between the polaritons and reads (see Supplementary Materials for more details)
\begin{align}\label{eq:main}
&\chi_{x+1}+\chi_{x-1}-2\chi_x\\ \nonumber
&+\left\{\tfrac{\omega_0+\omega_j-2\eps}{2\varphi \Gamma_0}+\tfrac{4}{Nk_j^2}\cos^2[ k_j (x-\tfrac12)]\right\}^{-1}
 \chi_x=0\:,
\end{align}
where $\omega_j\approx \omega_0 -2\varphi \Gamma_0/k_j^2$ is the real part of the eigenfrequency of the single-polaritonic state $\psi^{(j)}$ and $\varphi = \omega_0d/c$.
Equation~\eqref{eq:main} describes a motion of a particle on a lattice in an external potential of a standing wave  with the period $N/j$. It has a striking similarity to the  Harper equation for  an electron moving in a square lattice subjected to the perpendicular magnetic field~\cite{Harper1955}:
\begin{equation}\label{eq:Harper}
\chi_{x+1}+\chi_{x-1}+2\cos (2\pi x\alpha-k_y)\chi_x=\eps \chi_x\:.
\end{equation}
Here, $\alpha$ is the magnetic flux through the unit cell and $k_y$ is the wave vector in the perpendicular direction.
For small magnetic fields $\alpha \ll 1$, the energy spectrum of Eq.~\eqref{eq:Harper} is a ladder of degenerate Landau levels for electrons moving along quantized cyclotron orbits. In the finite structure, the edge states of topological nature arise in the gaps between the Landau levels. Such states correspond to electrons moving along skipping orbits at the structure edge, and are the origin for the quantum Hall effect~\cite{TKNN1982}.

In our system, the ratio $j/N$ of the cluster index to the total number of atoms in Eq.~\eqref{eq:main} plays the same role  as  the magnetic field flux in Eq.~\eqref{eq:Harper}. The spectrum also consists of degenerate Landau levels and the topological edge states in the gaps between them, see Fig.~\ref{fig:3}a.  The result of exact numerical diagonalization of the two-polariton Hamiltonian Eq.~\eqref{eq:H} [bold symbols in Fig.~\ref{fig:3}a] agrees quantitatively with  the solution of ~Eq.~\eqref{eq:main} [open symbols in Fig.~\ref{fig:3}a]. Thus, the states Eq.~\eqref{eq:psi} acquire  a peculiar internal structure, with nontrivial topology induced  by interaction for each of the two indistinguishable polaritons.
% \red{More details on the topological properties of Eq.~\eqref{eq:main} and the Chern numbers of the different allowed bands, separated by the band gaps are given in the Supplementary Information. }

The energy spectrum of the Harper Eq.~\eqref{eq:Harper}  becomes very rich when the magnetic flux $\alpha$ increases. The Landau levels split and transform into  a celebrated Hofstadter butterfly~\cite{Hofstadter1976}, shown  also in Supplementary Fig.~S2. The butterfly  has a self-similar structure with $q$ allowed energy bands at  the rational fluxes $\alpha=p/q$~\cite{Azbel1964} and a Cantor-set spectrum for irrational fluxes. Even though   in our case the effective magnetic flux $j/N$ is  rational, we can  still extract an analogue of the Hofstadter butterfly from the two-polariton  spectrum in Figs.~\ref{fig:2}{e--g}.
We separate the  clusters in Figs.~\ref{fig:2}{e--h} formed by different standing waves (i.e. different effective magnetic fields) and align  them horizontally, the details are presented in Methods and Supplementary Fig.~S3. The resulting butterfly is shown in Fig.~\ref{fig:3}b and it qualitatively resembles the Hofstadter butterfly [Fig.~S2].

 In accordance with Fig.~\ref{fig:3}a and Fig.~\ref{fig:2}, for small magnetic fluxes $j/N$ the butterfly in Fig.~\ref{fig:3}b features distinct Landau  levels with edge states in the gaps between them.  These edge states correspond to red points  in  Fig.~\ref{fig:3}b. At high magnetic fields the Landau levels split, but the spectrum still retains a surprisingly  delicate structure.

 %%%%%%%%%%%%%%
\begin{figure*}[t]
\includegraphics[width=\textwidth]{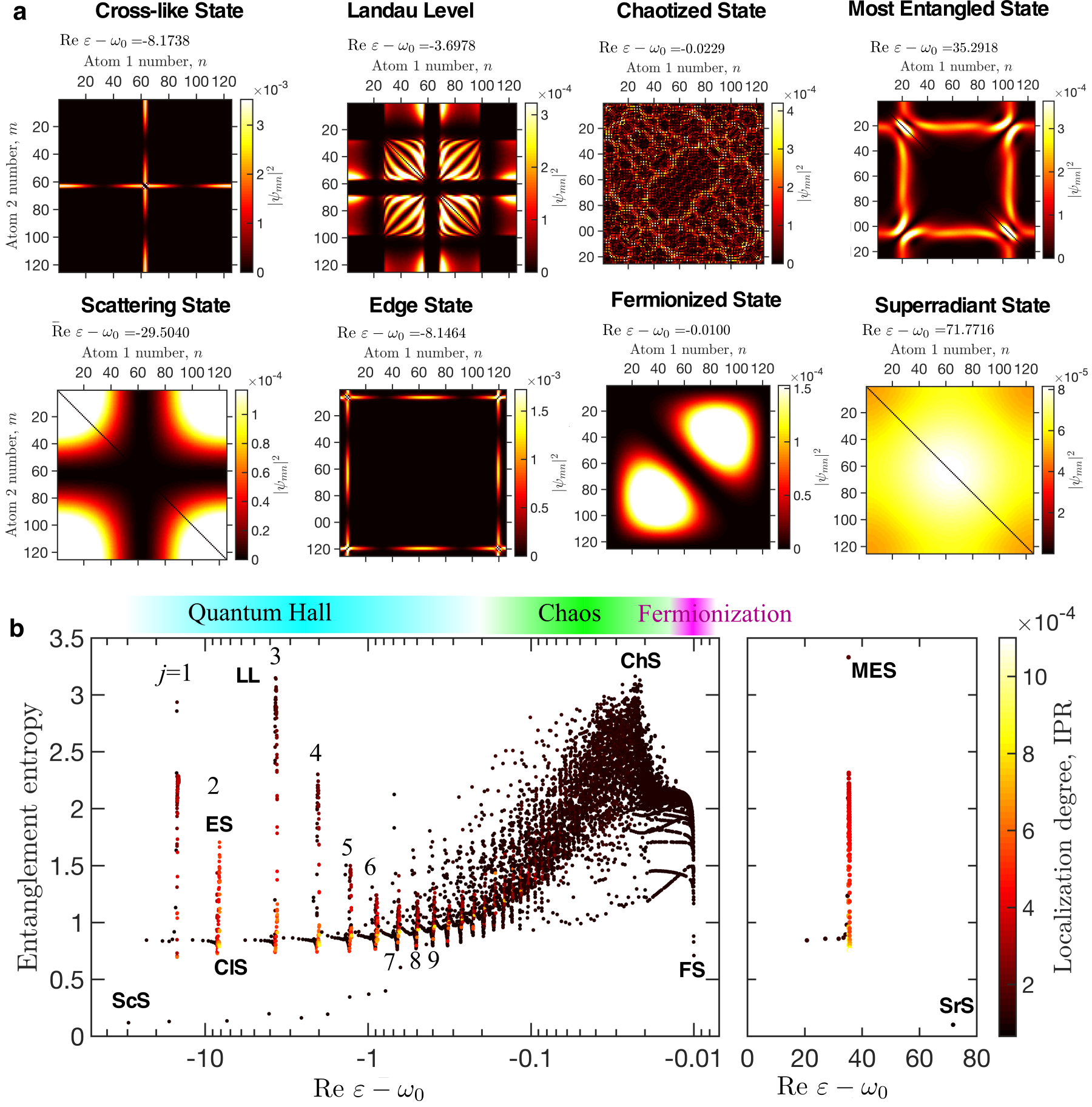}
\caption{{\bf Diversity of two-polariton states.}
{\bf (a)} Characteristic  wavefunctions for different types of two-polariton states, indicated in (b) by abbreviations.
{\bf (b)} Entanglement entropy depending on the state energy. Color shows the inverse participation ratio that characterizes the localization degree.
Left and right panels correspond to the states with Re~$\varepsilon<\omega_0$ and Re~$\varepsilon>\omega_0$, respectively.
Standing wave numbers $j$ are indicated near  the energy clusters.
 Calculation has been performed for $N=125$, $\omega_0d/c=0.02$. Energy is measured in the units of $\Gamma_0$.
}\label{fig:4}
\end{figure*}
%%%%%%%%%%%%%%
 %%%%%%%%%%%%%%%%%%%%%%%%%%%%%%%%%%%%%%%%%%%%%%%%%
 \section*{Polariton-polariton entanglement}
 %%%%%%%%%%%%%%%%%%%%%%%%%%%%%%%%%%%%%%%%%%%%%%%%%
The  internal structure of   the two-polariton states is represented by their entanglement entropy~
\cite{Eisert2010},  $S=-\sum\lambda_\nu \ln \lambda_\nu$, obtained from  the Schmidt expansion
$\psi_{nm}=\sum_{\nu=1}^N \sqrt{\lambda_\nu} \psi^\nu_n \psi^\nu_m\:,$  $(\sum \lambda_\nu=1)$.
 The result, presented   in Fig.~\ref{fig:4}b, demonstrates a rich variety of  eigenstates with different localization degrees, indicated by the color. Characteristic  examples of wave functions are shown in Fig.~\ref{fig:4}a. The  entropy of entanglement tells  us the number of distinct single-particle states in a given two-body state, so it  is low for the scattering states, where two polaritons are quasi-independent. The topological states  Eq.~\eqref{eq:psi} also have an intrinsically low entropy,
being just a product of a standing wave and a  localized  or an edge state.  However, the states Eq.~\eqref{eq:psi} can mix with each other resulting in larger entanglement entropy. This  entangled mixing becomes especially prominent
for the Landau level  states, cf. points LL, ClS and ES in Fig.~\ref{fig:4}b. When the real part of energy approaches $\omega_0$ from the negative side, the mixing between different standing waves increases since the spectrum gets denser,  and the states become chaotic-like, see also the top right corner of Fig.~\ref{fig:1}d. At  very small negative energies $\eps-\omega_0\sim-\varphi \Gamma_0$ the single-particle dispersion  changes from $\eps\propto -1/k^2$ to $\eps\propto -(k-\pi/d)^2$ because both polaritons get closer to the Brillouin zone edge,  and the fermionic correlations~\cite{Molmer2019} emerge from  chaos. The dense cluster of two-polariton   states in the right panel of Fig.~\ref{fig:4}b, where $\Re \varepsilon>0$, is formed by the interaction with the  quasi-superradiant mode with $\Re \eps-\omega_0\approx 71\Gamma_0$ in Fig.~\ref{fig:2}a. It is  one of the states of this cluster that has the entanglement entropy even higher than that of the chaotic-like states, see the point MES at $\Re \eps-\omega_0\approx 35 \Gamma_0$.

In summary, we have discovered a novel interaction-induced internal topological order for the two-polariton states in a light-coupled atomic array.
We have revealed that the underlying Dicke-type model demonstrates an incredible  diversity of quantum states with different topologies, lifetimes,  and entanglement in a strikingly simple system.  While its importance  for quantum optics~\cite{Roy2017,KimbleRMP2018} is already well understood, it  could rightfully take its place  also in the many-body physics, along with such celebrated examples  as a Heisenberg model, Bose-Hubbard model, or a Luttinger liquid. The waveguide-mediated long-ranged couplings are quite uncharacteristic for  traditional quantum systems and  there is much more to expect. For example, we have focused here only on the regime of extremely subwavelength distances between the atoms, where two-polariton bound states~\cite{Zhang2019arXiv,Poddubny2019quasiflat} play no role. Polariton-polariton  interactions could be even more interesting in  the Bragg-spaced lattices, where the non-Markovian effects are drastically enhanced~\cite{Ivchenko1994,Hubner1999,Goldberg2009}. The  ultra-strong coupling regime \cite{FriskKockum2019} is also unexamined  for the quantum waveguides to the best of our knowledge. On the more practical side, it is promising to explore  recently proposed  high-quality states \cite{Koshelev2020} in the many-body domain to increase the quantum coherence.   The  waveguide-based setups could be used to route and manipulate  signals~\cite{Leung2012}, propagating on future quantum chips~\cite{Arute2019} and  our results open new possibilities to engineer the quantum entanglement.

 \let\oldaddcontentsline\addcontentsline% Store \addcontentsline
\renewcommand{\addcontentsline}[3]{}% Make \addcontentsline a no-op

%%%%%%%%%%%%%%%%%%%%%%%%%%%%%%%%%%%%
{\bf Author contributions.} ANP  and AVP conceived the idea and developed  an analytical model.
JZ, ANP, NAO and YKe  performed the numerical calculations. ANP, CL, and YSK supervised the project.
All authors contributed to discussion of the results and writing the manuscript.

\acknowledgements{This work was supported by the Australian Research Council. J.Z was supported by Australian Government Research Training Program (RTP) Scholarship. C.L. was supported by the National Natural Science Foundation of China (NNSFC) (grants No. 11874434 and No. 11574405). Y.K. was partially supported by the Office of China Postdoctoral Council (grant No. 20180052), the National Natural Science Foundation of China (grant No. 11904419), and the Australian Research Council (DP200101168). A.V.P. acknowledges a support from the Russian Science Foundation (Project No. 19-72-00080). A.N.P. has been partially supported by the Russian President Grant No. MD-243.2020.2.  N.A.O. has been partially supported by the Foundation for the Advancement of Theoretical Physics and Mathematics ``BASIS.''  }

\setcounter{equation}{0}
\renewcommand{\thefigure}{M\arabic{figure}}
\renewcommand{\thesection}{M\Roman{section}}
\renewcommand{\thesection}{M\arabic{section}}
\renewcommand{\theequation}{M\arabic{equation}}
\section*{Methods}
Calculation of the energy spectrum of the Hamiltonian Eq.~\eqref{eq:H}, shown in Fig.~\ref{fig:2}, is relatively straightforward. The spectrum is found  by standard linear algebra techniques, see also Supplementary Materials.
However, it is more challenging to extract the butterfly spectrum in Fig.~\ref{fig:3}b from    the spectrum in Fig.~\ref{fig:2}e.
This task requires  careful separation of the clusters corresponding to different single-polariton states.
We start  by performing the Schmidt decomposition of the two-polariton state
\begin{equation}\label{eq:psi1b}
\psi_{xy}=\sum\limits_{\nu=1}^N \sqrt{\lambda_\nu} \psi^\nu_x \psi^\nu_y\:
\end{equation}
for all the states that have $\Re \varepsilon<\omega_0$.
Our analysis of the Schmidt decomposition confirms, that for most of the states it is governed  by two largest singular values $\lambda_1$ and $\lambda_2$, that have close absolute values. Keeping only these two terms, we obtain  new linear combinations of the wave functions $ \psi^1_x$ and $ \psi^2_y$ as
$u^{\pm}_n=\lambda_1^{-1/4}\psi^1_n\pm \lambda_2^{-1/4}\psi^{2}_n$\:.
After that, the two-polariton state can be approximately presented as
$\psi_{xy}\propto u^{+}_{x} u^{-}_{y}+u^{-}_{x} u^{+}_{y}\:.$
Next, we select one of the two states $u^{+}_{x}$,  $u^{-}_{y}$ that has  lower inverse participation ratio,
$\sum |u_x|^4/[\sum |u_x|^2]^2$\:, which means it is less localized in space. We designate this state as $u^{(\rm free)}$
and the more localized  one as $u^{(\rm loc)}$,
 perform the discrete Fourier transform
\begin{equation}
u^{(\rm free)}(k)=\sum\limits_{x=1}^N\e^{-\rmi kx}u^{(\rm free)}_x
\end{equation}
and calculate the wave vector $k_{\rm max}$, corresponding to the maximum of the Fourier decomposition. The number of the cluster can be then determined from the quantization rule
\begin{equation}\label{eq:cluster1}
j\approx \left[\frac{k_{\rm max}N}{\pi}\right],
\end{equation}
where square brackets indicate the rounding to the nearest integer.
In order to improve the precision in Eq.~\eqref{eq:cluster1} for large $j$, we also characterize the vectors $\chi$ by their mirror symmetry.
Then we apply Eq.~\eqref{eq:cluster1} separately for odd and even states with odd and even $j$, respectively. The results of Fourier transform  for $N=200$ atoms  are shown in Figs.~S3, S4 of the Supplementary Materials. Except for very large $j$ close to $N$, the spectrum is clearly separated into well-defined steps of alternating parity. Each step is assigned to a different cluster of eigenvalues. Next,
we align the clusters with respect to each other. This is done  by subtracting the energy with the largest (smallest negative) real part from the energies of the states of each cluster, $\eps_{j,\rm max}$. In order to keep the points with the  highest energy on the semilogarithmic plot after this subtraction, we also  add a small value of $1.1\times 10^{-4}\Gamma_0$ to all the energies.
 The result is the butterfly spectrum, shown in Fig.~\ref{fig:3}b.


%%%%%%%%%%%%%%
%\nocite{apsrev41Control}
%\bibliographystyle{apsrev4}
%\bibliography{titleon,hall}
%%%\bibliography{hall}

%%%%%%%%%%%%%%
\let\addcontentsline\oldaddcontentsline% Restore \addcontentsline

\newpage

\setcounter{figure}{0}
\setcounter{section}{0}
\setcounter{equation}{0}
\renewcommand{\thefigure}{S\arabic{figure}}
\renewcommand{\thesection}{S\Roman{section}}
\renewcommand{\thesection}{S\arabic{section}}
\renewcommand{\theequation}{S\arabic{equation}}

\begin{center}
\textbf{\Large Supplementary Materials}
\end{center}

\tableofcontents
%%%%%%%%%%%%%%%%%%%%%%%%%%%%%%%%%%%%%%%
\section{Analytical model for polariton-polariton interactions}
%%%%%%%%%%%%%%%%%%%%%%%%%%%%%%%%%%%%%%%
In this section, we start from the Hamiltonian (1) in the main text
\cite{Ke2019,Albrecht2019}
 \begin{align}\label{eq:HM}
 &H=\sum_n\omega_0 \sigma_n^\dag\sigma_n^{\vphantom{\dag}}+\sum\limits_{n,m}
 \sigma_n^\dag\sigma_m^{\vphantom{\dag}}\mathcal H_{nm}\:,\\
&\mathcal H_{nm}= -\rmi\Gamma_0 \e^{\rmi \varphi|n-m|},\quad \varphi=\frac{\omega_0d}{c}\:,
  \end{align}
and proceed to derive Eq.~(3) that describes an interaction of two polaritons.
    Substituting the  ansatz $|\Psi\rangle=\sum\psi_{mn} \sigma_n^\dag \sigma_m^\dag |0\rangle$ into the Schr\"odinger equation
$ H \Psi =2\eps\Psi$ we obtain   the two-polariton Schr\"odinger equation in the form~\cite{Ke2019,Zhong2020}
\begin{multline}\label{eq:S1}
\mathcal H_{mn'}\psi_{n'n}+\psi_{mn'}\mathcal H_{n'n}-2\delta_{mn}\mathcal H_{nn'}\psi_{n'n}\\=2(\eps-\omega_0) \psi_{mn}\:,
\end{multline}
or,
in a matrix form,
\begin{equation}\label{eq:S1b}
\mathcal H\psi+\psi \mathcal H-2\:{\rm diag\:}[{\rm diag\:} \mathcal H\psi]=2(\eps-\omega_0) \psi\:.
\end{equation}
Here, the  ``diag'' operator transforms a given matrix to the column-vector filled by the diagonal entries of this matrix, and vice versa.

This system is readily solved numerically after the  wavefunction $\psi$
is rewritten  in the basis
of $N(N-1)/2$ localized states  of the type
\[
[\widetilde\psi]_{mn}=[\widetilde\psi]_{nm}=\frac{1}{\sqrt{2}},\quad n\ne m\:.
\]

Our next goal is to go beyond Refs.~\cite{Ke2019,Zhong2020,Poddubny2019quasiflat} and obtain  Eq.~(3). To this end, we notice that
\cite{Zhong2020,Poddubny2019quasiflat}
\begin{equation}
K\equiv \mathcal H^{-1}\approx \frac{1}{2\varphi \Gamma_0}\partial^2,\text{ where } \partial^{2}\equiv \left(\begin{smallmatrix}
 -1&1&0&\ldots\\
 1&-2&1&\ldots\\
  &&\ddots&\\
  \ldots&1&-2&1\\
 \ldots  &0& 1&-1
 \end{smallmatrix}\right)\:.\label{eq:ih}
\end{equation}
Here, the matrix $\partial^{2}$ represents the one-dimensional discrete Laplacian (or the operator of discrete second-order derivative)
This means that for a  vector $\psi_n$ with a smooth dependence on $n$ one has
\[
[\partial^{2}\psi]_n=\psi_{n+1}+\psi_{n-1}-2\psi_n\approx\frac{\rmd^2 \psi_n}{\rmd n^2}\:.
\]
Thus,  for a short-period array with $\varphi\ll 1$ the operator $K$ reduces to the second derivative operator. The inverted Hamiltonian $K$ Eq.~\eqref{eq:ih} is a sparse matrix with only nearest-neighbor couplings. This fact inspires us to perform the transformation
\begin{equation}\label{eq:trans}
\psi=K \psi'K
\end{equation}
that means change of the basis to
\begin{equation}\label{eq:basis}
|x\rangle=\frac1{\sqrt{N}}\sum_{n=1}^N\e^{\rmi \omega_0d|x-n|/c}\sigma_n^\dag|0\rangle\:,
\text{where } x=1,2, \ldots N\:.
\end{equation}
This basis inherits the distribution of electric field emitted by a given atom. Indeed, $\exp(\rmi \omega_0d|x-n|/c)$ is just the photon Green function in one dimension. Since the  wave equations for electric field are local, the transformed two-polariton Schr\"odinger
equation will  be  local as well, i.e. it will involve  only sparse matrices. Substituting Eq.~\eqref{eq:trans}
into Eq.~\eqref{eq:S1b} we find \cite{Zhong2020,Poddubny2019quasiflat}
\begin{equation}
K\psi'+\psi'K-2\:{\rm diag\:}[{\rm diag\:} \psi' K]=2(\eps-\omega_0) K\psi'K\:.\label{eq:Sh3}
\end{equation}
Next, we look for  the solution of the transformed equation  \eqref{eq:Sh3} in the form
\begin{equation}\label{eq:ansatz}
\psi'_{xy} = \psi^{(j)}_{y} \chi_{x}+\psi^{(j)}_{x}\chi_{y}\:,\quad x,y=1\ldots N\:,
\end{equation}
corresponding to Eq.~(2) in the main text. Here, one of the two excitations is a single-particle eigenstate of the matrix $\mathcal H$ with the eigenfrequency $\omega_j$. Using the definition $K\equiv \mathcal H^{-1}$\:,e we find
\begin{equation}\label{eq:u}
  K\psi^{(j)}=\frac{1}{\omega_j} \psi^{(j)}\:.
\end{equation}
The state is normalized  as $ \sum_x [\psi_x^{(j)}]^2=1$. The normalization does not involve  complex conjugation, because the original matrix $\mathcal H$ is not Hermitian but symmetric. As such, its eigenvectors $\psi^{(j)}$ satisfy the non-conjugated orthogonality condition
\begin{equation}
\langle j|j'\rangle\equiv  \sum\limits_{x=1}^N \psi_x^{(j)}\psi_x^{(j')}=\delta_{jj'}\:.
\end{equation}
Due to the translational symmetry the vector $ \psi^{(j)}$ is just a standing wave~\cite{Ivchenko1991}:
\begin{equation}\label{eq:ansatz1}
 \psi^{(j)}_x\approx \sqrt{\frac{2}{N}}\cos \frac{\pi j(x-1/2)}{N}\:.
\end{equation}
We note, that the ansatz \eqref{eq:ansatz} and \eqref{eq:ansatz1} where  the eigenstate $\psi^{(j)}$ does not take  into account the interaction effects,  works only for the
transformed Schr\"odinger equation \eqref{eq:Sh3}. This ansatz  does not adequately describe the solutions of the  original equation \eqref{eq:S1b} because  the wavefunction $\psi'$ does not turn to zero for $n=m$.

%%%%%%%%%%%%%%
\begin{figure*}[t]
\includegraphics[width=0.8\textwidth]{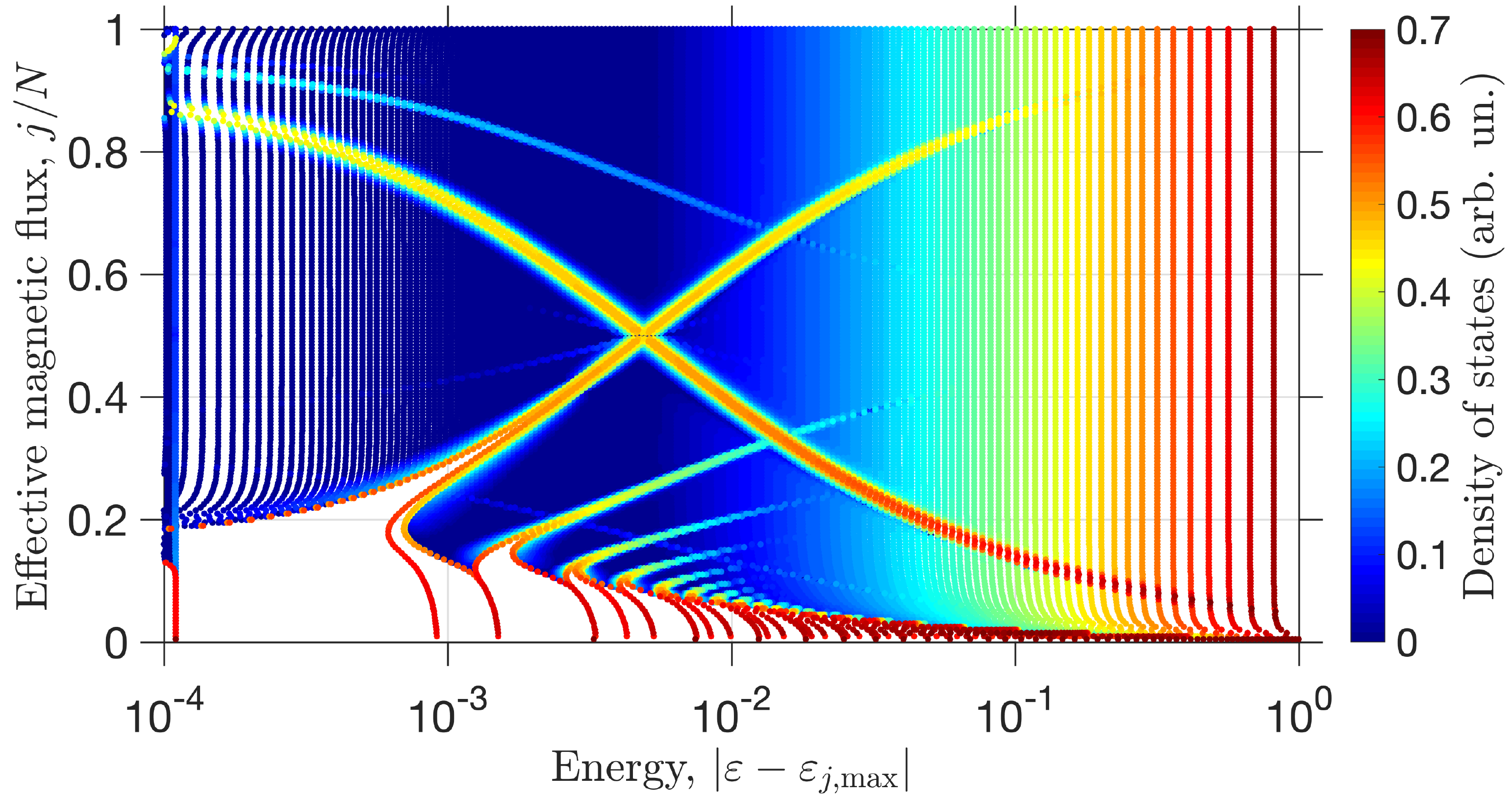}
\caption{Butterfly spectrum calculated from Eq.~\eqref{eq:main2} for $j=200$, $\varphi=0.02$\:.
Energy is measured in the units of $\Gamma_0$.
}\label{fig:anbut}
\end{figure*}
%%%%%%%%%%%%%%

Substituting Eq.~\eqref{eq:ansatz} into Eq.~\eqref{eq:Sh3}, we obtain
\begin{multline}
\frac{\psi_n^{(j)}}{\omega_j} \chi_m+(K\chi)_n \psi^{(j)}_m+
\frac{\psi_m^{(j)}}{\omega_j} \chi_n+(K\chi)_m \psi^{(j)}_n\\-
2\delta_{mn}\left[\frac{\psi^{(j)}_n}{\omega_j} \chi_n+(K\chi)_n \psi^{(j)}_n\right]\\=\frac{2(\eps-\omega_0)}{\omega_j} \left[ \psi^{(j)}_n (K\chi)_m+\psi^{(j)}_m (K\chi)_n\right]\:.
\end{multline}
Next, we multiply this equation  by $\psi^{(j)}_m$ and sum over $m$:
\begin{multline}\label{eq:sum1}
\frac{1}{\omega_j}\chi_n+(K\chi)_n+ \psi^{(j)}_n \psi^{(j)}_m (K\chi)_m+\frac{\psi^{(j)}_n}{\omega_j}\chi_m \psi^{(j)}_m
\\-\frac{2}{\omega_j} \psi^{(j),2}_n \chi_n-2\psi^{(j),2}_n (K\chi)_n\\=\frac{2(\eps-\omega_0)}{\omega_j} \left[ \psi^{(j)}_n \psi^{(j)}_m(K\chi)_m+ (K\chi)_n\right]\:.
\end{multline}
For the sake of brevity, the summation over the dummy index $m$ is assumed but not indicated explicitly.
Using the fact that $\psi^{(j)}$ is the eigenstate of $K$, we simplify Eq.~\eqref{eq:sum1} to
\begin{multline}\label{eq:sum2}
\left[\frac{\chi_n}{\omega_j}+K_{nm}\chi_m+\frac{2}{\omega_j} \psi^{(j)}_n \psi^{(j)}_m \chi_m\right]\\-2 \psi^{(j),2}_n \left[\frac1{\omega_j} \chi_n+K_{nm}\chi_m\right]\\=\frac{2(\eps-\omega_0)}{\omega_j}\left(K_{nm}\chi_m+\frac{1}{\omega_j}  \psi^{(j)}_n \psi^{(j)}_m \chi_m\right)\:.
\end{multline}
We are going to consider strongly localized eigenstates that are orthogonal to the standing wave $\psi^{(j)}$.
Thus, the terms $\propto \psi^{(j)}_m \chi_m$ in Eq.~\eqref{eq:sum2} can be omitted to find our main result:
\begin{multline}\label{eq:main2}
\left[\frac{1}{\omega_j}+K\right]\chi-\frac{2}{\omega_j}\diag [\psi^{(j),2}]\chi-\diag [(\psi^{(j),2})K]\chi\\=\frac{2(\eps-\omega_0)}{\omega_j} K \chi\:.
\end{multline}
Moreover, for relatively small $j$ the function $\chi$ changes with $n$ much faster than $\psi^{(j)}$. Hence,
 the term, proportional to $\diag [(\psi^{(j),2})K]\chi$, is larger  than the terms
 $\diag [\psi^{(j),2}]\chi$. Neglecting the terms~$\propto \psi^{(j),2}_n \chi_n$, we  obtain
\begin{equation}\label{eq:main2}
\left[\frac{1}{\omega_j}+K\right]\chi-2\diag [(\psi^{(j),2})K]\chi=\frac{2(\eps-\omega_0)}{\omega_j} K \chi\:.
\end{equation}
Taking Eq.~\eqref{eq:ih}  into account, we get Eq.~(3) from the main text, i.e.
\begin{align}\label{eq:mainb}
&\chi_{x+1}+\chi_{x-1}-2\chi_x\\ \nonumber
&+\left\{\tfrac{\omega_0+\omega_j-2\eps}{2\varphi \Gamma_0}+\tfrac{4}{Nk_j^2}\cos^2[ k_j (x-\tfrac12)]\right\}^{-1}
 \chi_x=0\:.
\end{align}
The butterfly spectrum, calculated from Eq.~\eqref{eq:main2}, is shown in Fig.~\ref{fig:anbut}.
In the calculation we have neglected the imaginary part of $\omega_j$, replaced the exact operator $K=\mathcal H^{-1}$
by its approximation  Eq.~\eqref{eq:ih}  and used the analytical expression
Eq.~\eqref{eq:ansatz1} for the wavefunction $\psi^{(j)}$.
In order to better resolve the small band gaps at large $j$ we have colored the points by the density of states, with red meaning more dense spectrum. Namely, the color scale corresponds to the logarithm of the fourth derivative of the spectrum $\varepsilon_\nu$, $\nu=1\ldots N$, obtained separately for each value of $j$.
 For low $j$, the result in  Fig.~\ref{fig:anbut}  is generally similar to the numerically obtained butterfly, shown in Fig.~3b of the main text. To our surprise, the exact numerical spectrum in Fig.~3b of the main text is actually richer than the semi-analytical one in Fig.~\ref{fig:anbut}.
%%%%%%%%%%%%%%
\begin{figure}[t!]
\includegraphics[width=0.45\textwidth]{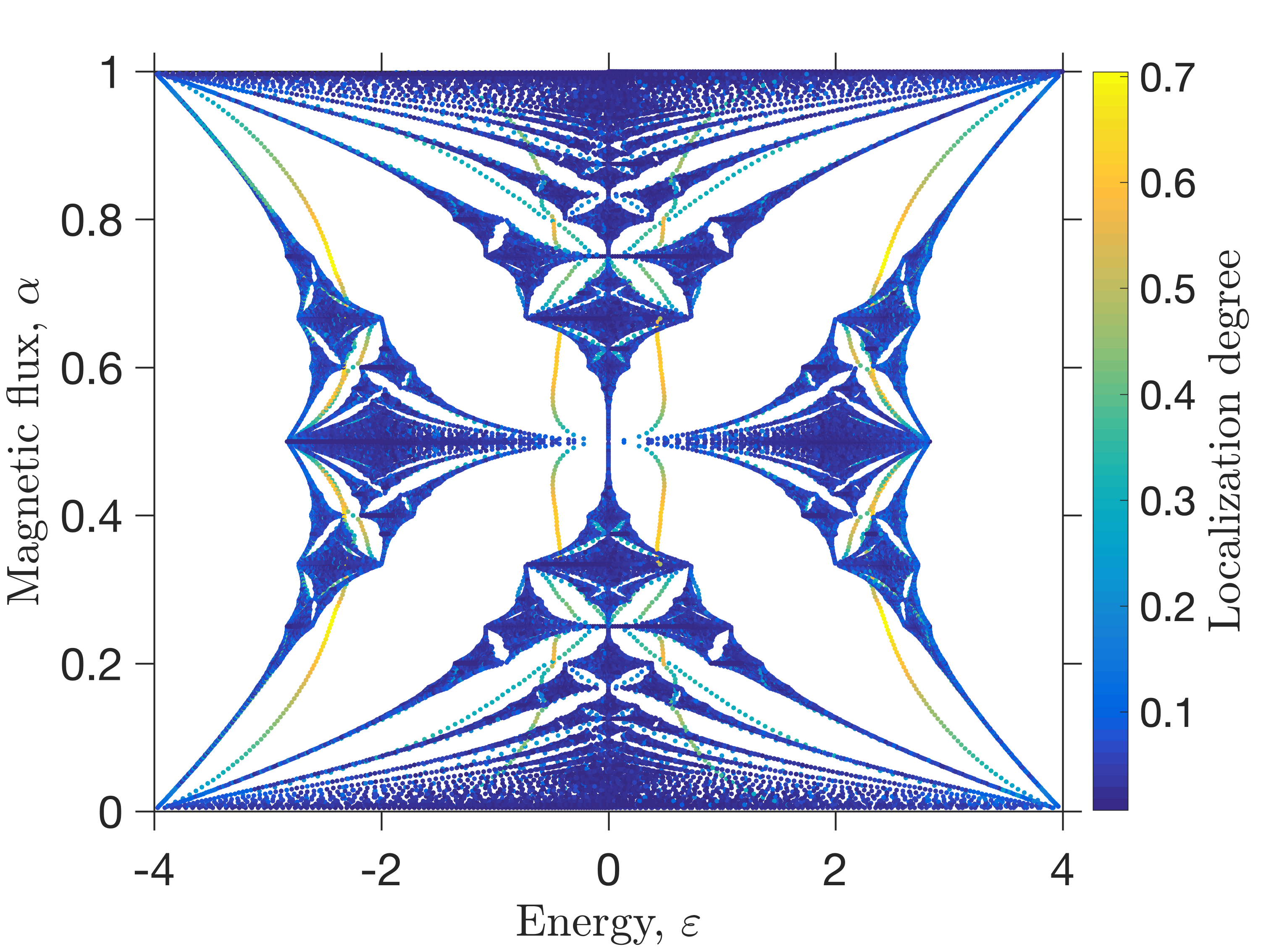}
\caption{Hofstadter butterfly obtained from solution of Eq.~\eqref{eq:Harper1}. Calculation has been performed for an array with $N=200$ sites, $k_y=0$, open boundary conditions, and the flux $\alpha$ changing from $0$ to $1$ with the step $1/(4N)$. Color of the points corresponds to their inverse participation ratio (IPR), higher IPR corresponds to topological edge states.}\label{fig:OrigBut}
\end{figure}
 %%%%%%%%%%%%%%

For comparison, we also present in Fig.~\ref{fig:OrigBut} the original Hofstadter butterfly~\cite{Hofstadter1976}. It is obtained by
studying the dependence of the spectrum of  the Harper equation~\cite{Harper1955} (Eq.~(4) in the main text)
on the magnetic flux $\alpha$
\begin{equation}\label{eq:Harper1}
\chi_{x+1}+\chi_{x-1}+2\cos (2\pi x\alpha-k_y)\psi_x=\eps \chi_x\:.
\end{equation}
Three butterfly spectra, calculated  from full two-polariton Hamiltonian Eq.~\eqref{eq:HM}, semi-analytical  model Eq.~\eqref{eq:mainb} and the Harper equation Eq.~\eqref{eq:Harper1}, are shown in Fig.~\ref{fig:all} (corresponding to Fig.~3b of the main text), Fig.~\ref{fig:anbut} and  Fig.~\ref{fig:OrigBut}, respectively. These  spectra have both similarities and differences. At low magnetic fields they all demonstrate distinct degenerate Landau levels separated by the band gaps with topological edge states. However, their behavior diverges at higher magnetic fields. The difference between
the semi-analytical butterfly in Fig.~\ref{fig:anbut} and the original Hofstadter butterfly in  Fig.~\ref{fig:OrigBut} could be attributed to the peculiar nature of Eq.~\eqref{eq:mainb}. Namely, in contrast to the conventional eigenvalue problem Eq.~\eqref{eq:Harper1} where the energy $\eps$ enters the term in the right-hand side, Eq.~\eqref{eq:mainb} contains the energy $\eps$ in the  denominator of the last term, i.e. it is a generalized eigenvalue problem. This is related to the long-ranged photon-mediated couplings between the atoms.
Namely, the price for the transformation from  the original Hamiltonian Eq.~\eqref{eq:S1} to the  model Eq.~\eqref{eq:Sh3} with nearest-neighbor couplings was the appearance of the $K$  operators in the right hand side of Eq.~\eqref{eq:Sh3}, i.e. the
Eq.~\eqref{eq:Sh3} became a generalized eigenvalue problem.  It is not fully clear at the moment, which of the conclusions derived from the original Harper model  Eq.~\eqref{eq:Harper1} should be valid for the generalized Aubry-Andr\'e-Harper model Eq.~\eqref{eq:main}.

As mentioned above, the spectrum of the full Hamiltonian Eq.~\eqref{eq:HM} is even richer than that of the generalized Aubry-Andr\'e-Harper model Eq.~\eqref{eq:main}. The reason for this might be the mixing between  the standing waves with different orders $j$ that is not accounted in the ansatz Eq.~\eqref{eq:ansatz1}.  This mixing becomes prominent for high values of $j$, as is also indicated by the calculation of the  entanglement entropy in Fig.~4 of the main text. It is important, that the butterfly has a fine structure, band gaps and edge states even for high values  of effective magnetic flux $j/N\sim 0.3$, see in particular the top state in the left column of Fig.~\ref{fig:all}.

%%%%%%%%%%%%%%%%%%%%%%%%%%%%%%%%%%%%%%%
\section{Fourier analysis of the eigenstates}
\begin{figure}[b!]
\includegraphics[width=0.45\textwidth]{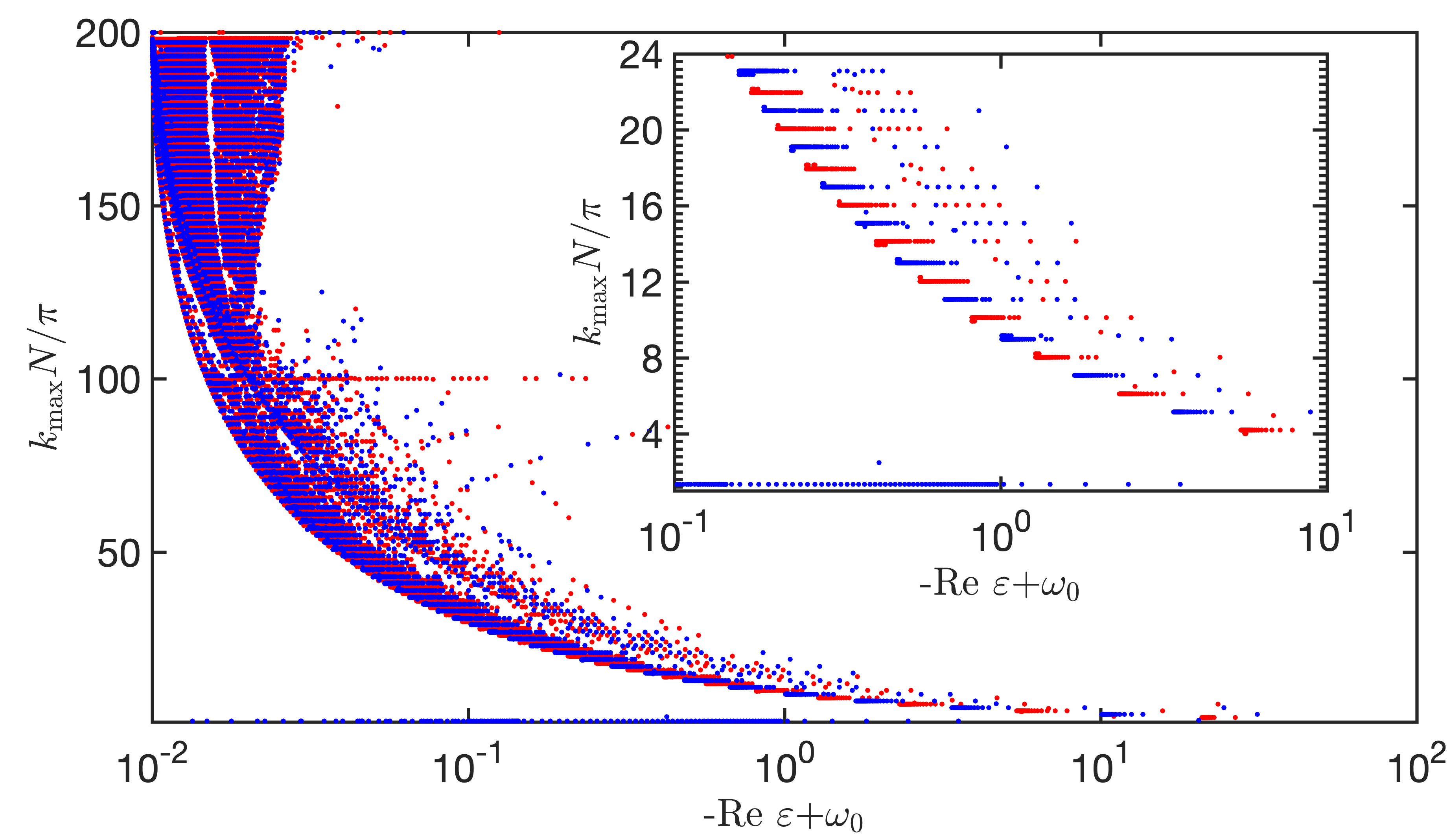}
\caption{Fourier analysis of the eigenstates for $N=200$ atoms, $\varphi=0.02$. Red and blue colors correspond to even and odd eigenstates, respectively. Inset shows the spectrum in a larger scale for small $k_{\rm max}$. Energy is measured in the units of $\Gamma_0$.}\label{fig:fourier}
\end{figure}

\begin{figure*}[b]
\includegraphics[width=0.8\textwidth]{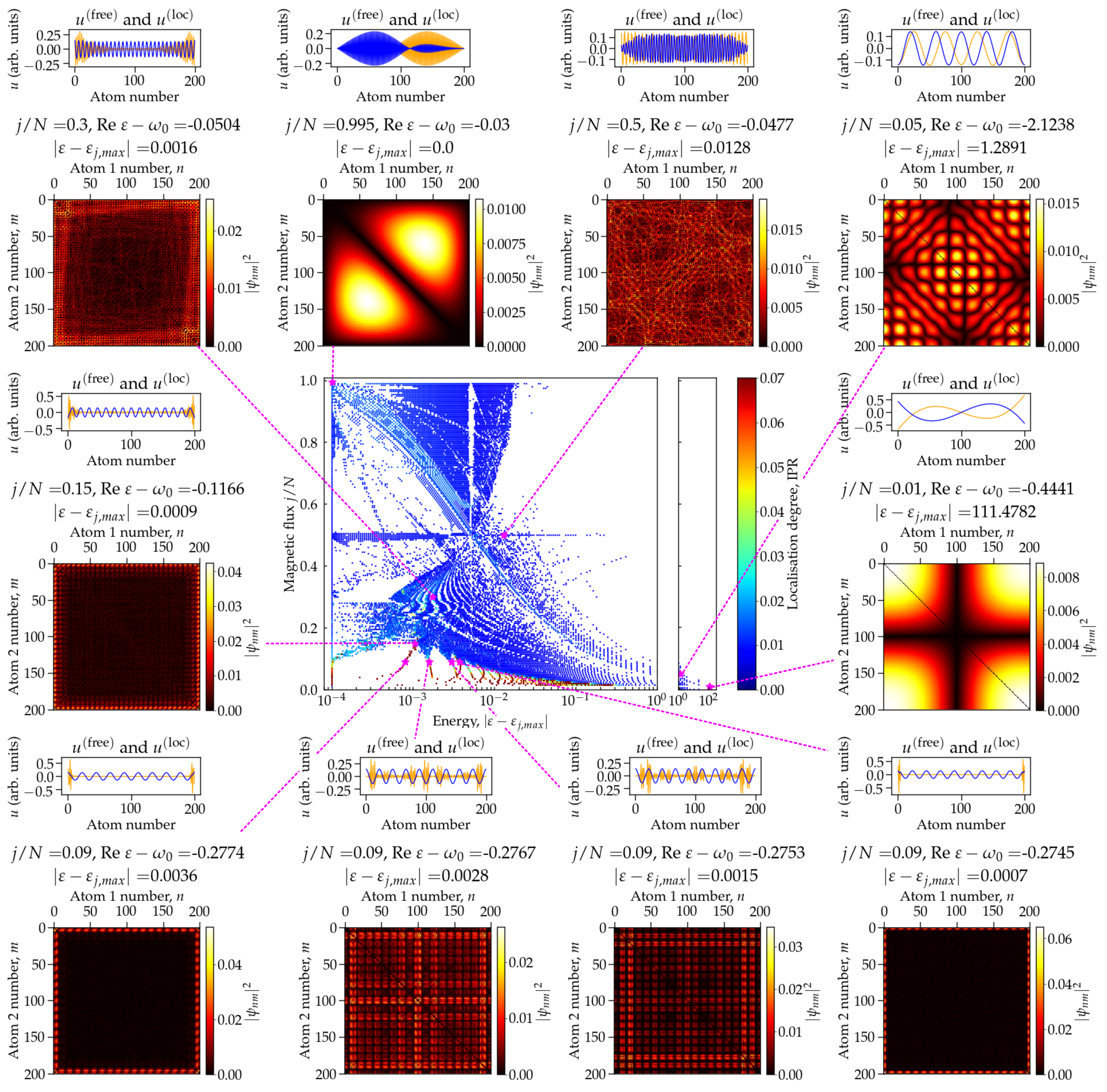}
\caption{Butterfly spectrum, corresponding to Fig.~3 from the main text, and several characteristic two-polariton states. For each state we show the whole two-polariton wave function and the real parts of the two vectors $u^{(\rm loc)}$ and $u^{(\rm free)}$, corresponding to leading terms in its singular value decomposition, see Methods.
Calculation has been performed for $N=200$ and $\varphi=0.02$, energy is measured in the units of $\Gamma_0$
}\label{fig:all}
\end{figure*}

%%%%%%%%%%%%%%%%%%%%%%%%%%%%%%%%%%%%%%%
We present in Fig.~\ref{fig:fourier} the result of the Fourier analysis of the eigenstates, described in the Methods section of the main text.  Figure~\ref{fig:all} shows the examples of singular value decomposition for several characteristic two-polariton states.

\end{document}